\documentclass{nature_meth}
\usepackage{amssymb,amsfonts,amsmath}
\usepackage{graphicx} 
\usepackage{gensymb}

 \usepackage{multirow}
\usepackage{epsfig}
\usepackage[figoff]{figcaps}
\usepackage{hyperref}
\hypersetup{
    colorlinks=true,
    linkcolor=blue,
    filecolor=magenta,
    urlcolor=cyan,
}

\usepackage{outline}
\usepackage{pmgraph}
\usepackage[normalem]{ulem}
\usepackage[utf8]{inputenc}
\usepackage{amssymb}
\usepackage{hyperref}
\usepackage{amsmath}
\usepackage{graphicx}
\usepackage{times}
\usepackage{xcolor}
\usepackage{xspace}
\usepackage[colorinlistoftodos]{todonotes} 
\usepackage{cite}
\usepackage{bm} 
\usepackage{url}

\usepackage{epstopdf}
\AppendGraphicsExtensions{.tiff}
\graphicspath{{fig/}} 

\usepackage{epsfig}
\usepackage{tikz}
\usetikzlibrary{spy}
\usepackage{algpseudocode}
\usepackage{algorithm}
\usepackage{mathrsfs}

\long\def\comment#1{} 

\newcommand{\beq}{\begin{equation}}
\newcommand{\eeq}{\end{equation}}
\newcommand{\beqa}{\begin{eqnarray}}
\newcommand{\eeqa}{\end{eqnarray}}

\usepackage{caption,setspace}
\renewcommand{\spacing}[1]{\renewcommand{\baselinestretch}{#1}\large\normalsize}
\spacing{2}

\usepackage{url}

\usepackage{graphicx}
\makeatletter
\let\saved@includegraphics\includegraphics
\AtBeginDocument{\let\includegraphics\saved@includegraphics}
\makeatother

\title{AI can evolve without labels: self-evolving vision transformer for chest X-ray diagnosis through knowledge distillation}
\author{Sangjoon Park$^{1}$, 
Gwanghyun Kim$^{1}$, 
Yujin Oh$^{1}$,
Joon Beom Seo$^{2}$, 
Sang Min Lee$^{2}$, 
Jin Hwan Kim$^{3}$, 
Sungjun Moon$^{4}$, 
Jae-Kwang Lim$^{5}$, 
Chang Min Park$^{6}$, 
and Jong Chul Ye$^{\dagger,1}$
}

\begin{document}
\setstretch{1.2}

\maketitle

\begin{affiliations}
\item Department of Bio and Brain Engineering, KAIST, Daejeon, Korea
\item Asan Medical Center, University of Ulsan College of Medicine, Seoul, South Korea
\item College of Medicine, Chungnam National Univerity, Daejeon, South Korea
\item College of Medicine, Yeungnam University, Daegu, South Korea
\item School of Medicine, Kyungpook National University, Daegu, South Korea
\item College of Medicine, Seoul National University, Seoul, South Korea
\item[] $^{\dagger}$Correspondence should be addressed to J.C.Y. (jong.ye@kaist.ac.kr)
\end{affiliations}
 
\vspace{-0.5cm}
\section*{Abstract}
\begin{abstract}
Although deep learning-based computer-aided diagnosis systems have recently achieved expert-level performance, developing a robust deep learning model requires large, high-quality data with manual annotation, which is expensive to obtain. This situation poses the problem that the chest x-rays collected annually in hospitals cannot be used due to the lack of manual labeling by experts, especially in deprived areas. 
To address this, here we present a novel deep learning framework that uses knowledge distillation through self-supervised learning and self-training, which shows that the performance of the original model trained with a small number of labels can be gradually improved with more unlabeled data.
Experimental results show that the proposed framework maintains impressive robustness against a real-world environment and has general applicability to several diagnostic tasks such as tuberculosis, pneumothorax, and COVID-19. Notably, we demonstrated that our model performs even better than those trained with the same amount of labeled data.
The proposed framework has a great potential for medical imaging, where plenty of data is accumulated every year, but ground truth annotations are expensive to obtain.
\end{abstract}

\clearpage
\setstretch{1.6}

\section*{Introduction}

{With the early success of deep learning for medical imaging \cite{gulshan2016development, de2018clinically, rajpurkar2017chexnet}, the application of artificial intelligence (AI) for the medical image has rapidly accelerated in recent years \cite{ting2018ai, giger2018machine, pesapane2018artificial}.
In particular, many deep learning based computer-aided diagnosis (CAD) softwares have been introduced into routine practice \cite{lakhani2017deep, pasa2019efficient, harris2019systematic, qin2021tuberculosis} for various imaging modalities such as chest X-ray (CXR). These deep learning-based AI models have demonstrated the potential to dramatically reduce the workload of clinicians  in a variety of contexts if used as an assistant, leveraging their power to handle a large corpus of data in parallel. The advantage can be maximized in resource-limited settings such as underdeveloped countries 
where various diseases such as tuberculosis prevail while the experts to provide the accurate diagnosis are scanty.}

 Most of the existing AI tools are based on the  convolutional neural network (CNN) models built with supervised learning, but collecting large and well-curated data with the ground truth annotation is rather difficult in the underprivileged areas where the amount of available data itself is abundant. In particular, although the size of data increases in number every year in these areas, the lack of ground truth annotation hinders the use of increasing number of data to improve the performance of AI models.

Given the limitation in label availability, an important line of machine learning research is  self-supervised and semi-supervised learning, which relies less on the corpus of labeled data. In general, the orthodoxy was that a model trained with a supervised learning approach is the upper bound of the performance. 
However, it was recently shown that the {\em self-training} with knowledge distillation between the teacher and noisy student, a type of semi-supervised learning approach, can substantially improve the robustness of the model to adversarial perturbations. The key idea of this method is to train two separated student and teacher models so that the student is trained with images with various forms of {noise} to meet the teacher's prediction with the same but {clean} image. Experimental results suggest  that the  knowledge distillation with enough noise can do better in various external validation settings than the traditional supervised model. In addition, a {recently developed} Vision transformer (ViT) \cite{dosovitskiy2020image} was successfully
utilized in a  method called the distillation with no label (DINO)~\cite{caron2021emerging} by exploiting 
the knowledge distillation between student and teacher via the local to global view correspondence for {\em self-supervised learning}. Besides achieving a new SOTA performance among self-supervised learning approaches, the powerful self-attention mechanism in ViT can segment objects without supervision, showing that the model is capable of a higher-level image understanding. {Inspired by  that both methods are based on the knowledge distillation between teacher and student}, here we suggest a ViT-based self-evolving framework for CXR diagnosis that can gradually improve the performance simply with an increasing amount of unlabeled data, amalgamating the distinct strengths of self-supervised learning and self-training through knowledge distillation.

{Our method, dubbed distillation for self-supervised and self-train learning (DISTL), can gradually improve the performance of the AI model   in various external validation settings
by maximally utilizing the common ground of knowledge distillation from self-supervised and self-training with the increasing amount of unlabeled data
(Fig.~\ref{fig1}a).}
Of note, it even outperforms the supervised model trained with the same amount of labeled data in the external validation. Furthermore,  the proposed self-evolving method has substantial robustness to the real-world data corruptions, and our model offers a more straightforward visualization of the model's attention to locating the lesion.
We argue that the distillation of knowledge through self-training and self-supervised learning, even without knowledge of the lesion, results in a high correlation of attention with the lesion, which may be the reason for the superior performance in diagnosis.

\subsection{Overview of the proposed framework}
\begin{figure}[!t]
	\centering
 \centerline{\epsfig{figure=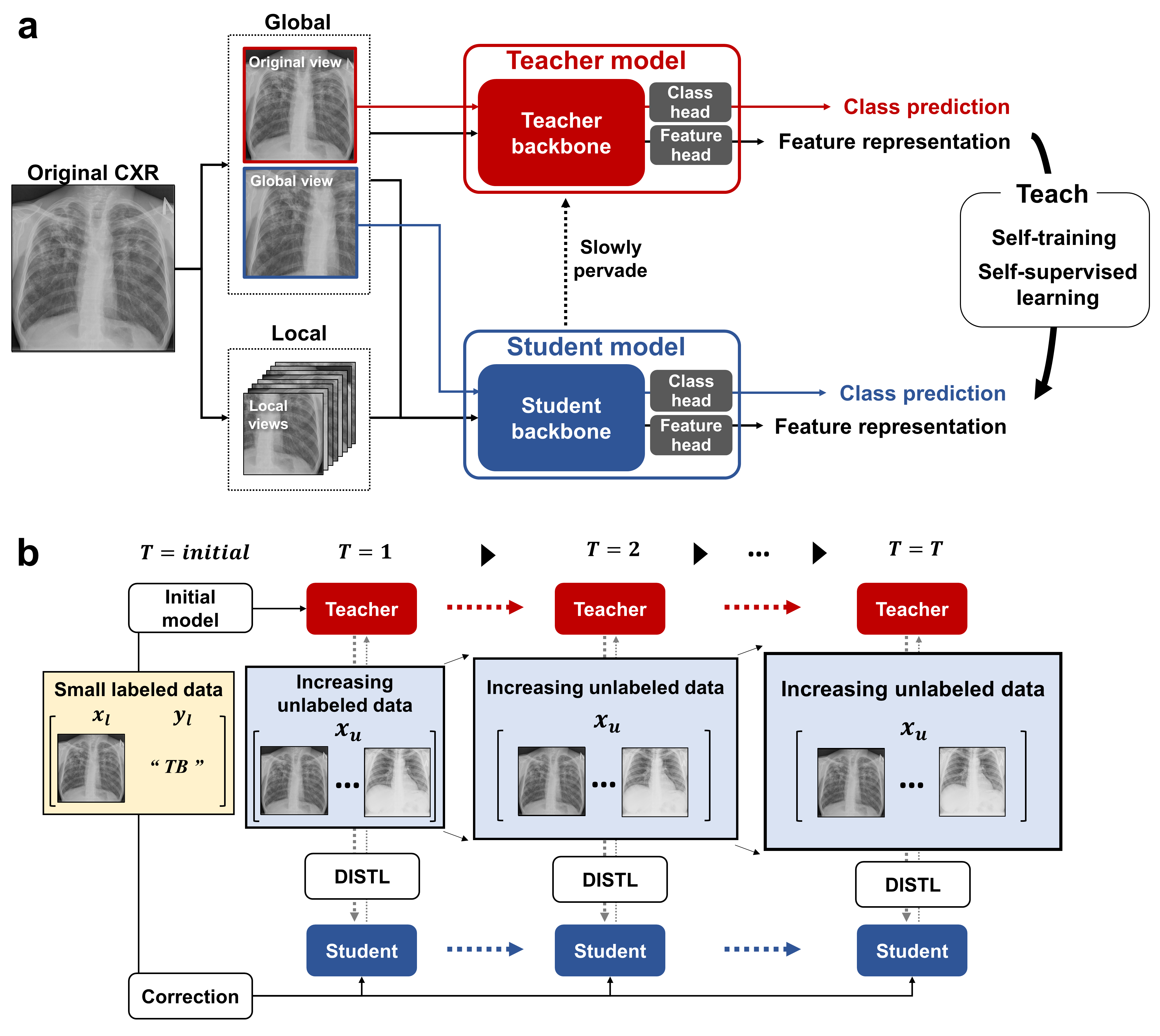, width=0.85\linewidth}}
	\caption{\bf\footnotesize 
Overview of the proposed framework for the self-evolving AI model. 
(a) The distillation for self-supervised and self-train learning (DISTL) method is composed of the two components, for \emph{self-supervision} and \emph{self-training}. 
(b) The initial model is trained with small labeled data. Then, using this model as the teacher, the student is trained with the DISTL method under an increasing amount of data over time $T$.}
	\label{fig1}
\end{figure}

As shown in Fig.~\ref{fig1}a,  to stably evolve our model performance leveraging unlabeled cases,  the two identical models, teacher, and student are utilized for distillation, encouraging the student model to match its \emph{noised} prediction obtained from a given CXR to the \emph{clean} prediction of teacher model obtained from the same CXR. 
However, unlike the previous noisy self-training approach, both self-supervision and self-training methods were leveraged in our method. Specifically, self-supervision plays a key role to incentivize the model to learn the task-agnostic semantic features of the CXR by having more shape bias to the image content (Supplementary Fig.~1), while the self-training enables the model to directly learn the task-specific information, for example,  the diagnosis of tuberculosis. To verify this hypothesis, we conducted ablation studies by removing each component, demonstrating these two components are imperative to attain the optimal performance (Supplementary Fig.~2).
The details of our algorithm and ablation studies can be found in Supplementary Material.

 In our method, to gradually evolve the model performances given the increasing unlabeled data accumulated over time (Fig.~\ref{fig1}b), the initial model was first built using supervised learning with the small labeled data. Then, we used this initial model as the teacher to train the student in large unlabeled data. In this process, the teacher is slowly co-distilled from the updated student. In addition, to prevent the student from being deteriorated by the imperfect estimation of the teacher, the {correction} with the initial small labeled data was done per the predefined steps. The updated model is then used as the starting point of the next-generation model, similar to the previous self-training approach with increasing time $T$ \cite{xie2020self}.

\begin{figure}[!t]
	\centering
 \centerline{\epsfig{figure=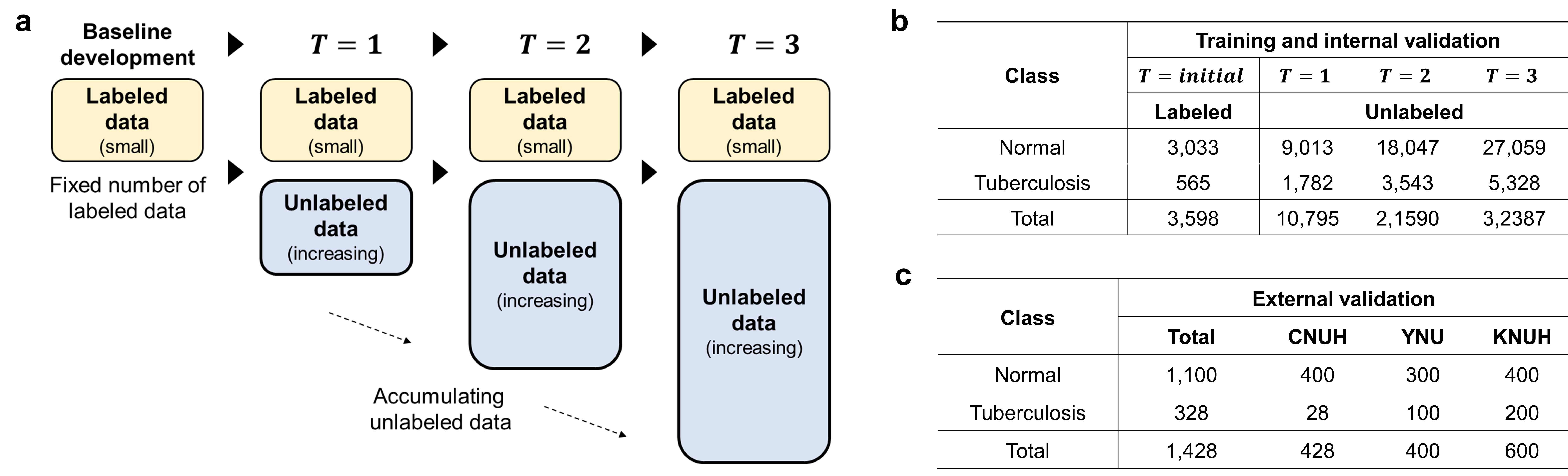, width=1.0\linewidth}}
	\caption{\bf\footnotesize 
(a) Simulation of clinical application of increasing data over time.
To simulate the increasing data over time $T$, we divided Chest X-rays (CXRs) into labeled and unlabeled subsets. Then, we gradually increased the amount of available unlabeled data, supposing the time $T$ goes.
(b) Details of data partitioning and classes used for training and internal validation.
(c) For external validation, data were deliberately collected in three hospitals.}
	\label{fig2}
\end{figure}

\section*{Results}

{We evaluated the proposed framework in three CXR tasks including the diagnosis of tuberculosis by using only a small corpus of labeled data for supervision and gradually increasing the amount of unlabeled data simulating the real-world data accumulation over time.} 

In particular, to confirm whether our AI model can gradually self-evolve in the data-abundant but the label-insufficient situation, 
 we set our main task as the diagnosis of tuberculosis (TB), as it is highly demanded in clinics after World Health Organization has identified the use of AI-based CAD for CXR screening of tuberculosis as a potential solution in resource-limited settings \cite{who2021global}.
{We collected the normal and tuberculosis CXRs from both the publicly available open-source and the institutional data sets for the model development and internal validation (Supplementary Fig.~3 and ``Details of datasets for diagnosis" section). After collection, a total of 35,985 CXRs were further divided into 3,598 labeled (10\% of total data) and 32,387 unlabeled subsets (90\% of total data). Next, assuming the situation in the clinic that the number of unlabeled cases increases as time goes, the unlabeled subset was further divided into three. Then, using these three folds, we increased the total amount of available unlabeled data to be 30\%, 60\%, 90\% of total data, supposing the time $T = 1, 2, 3$ goes as shown Fig.~\ref{fig2}a. During this process, the subset of labeled data remains fixed to the initial 3,598 CXRs (10\% of total data) (Fig.~\ref{fig2}b). The performances of the proposed self-evolving AI model at each time $T$ were evaluated in the external validation data collected and labeled by board-certified radiologists in three different hospitals (Chonnam National University Hospital [CNUH], Yeungnam University Hospital [YNU], and Kyungpook National University Hospital [KNUH]), to validate the generalization capability for different devices and image acquisition settings (Fig.~\ref{fig2}c).}

\begin{figure}[!t]
	\centering
 \centerline{\epsfig{figure=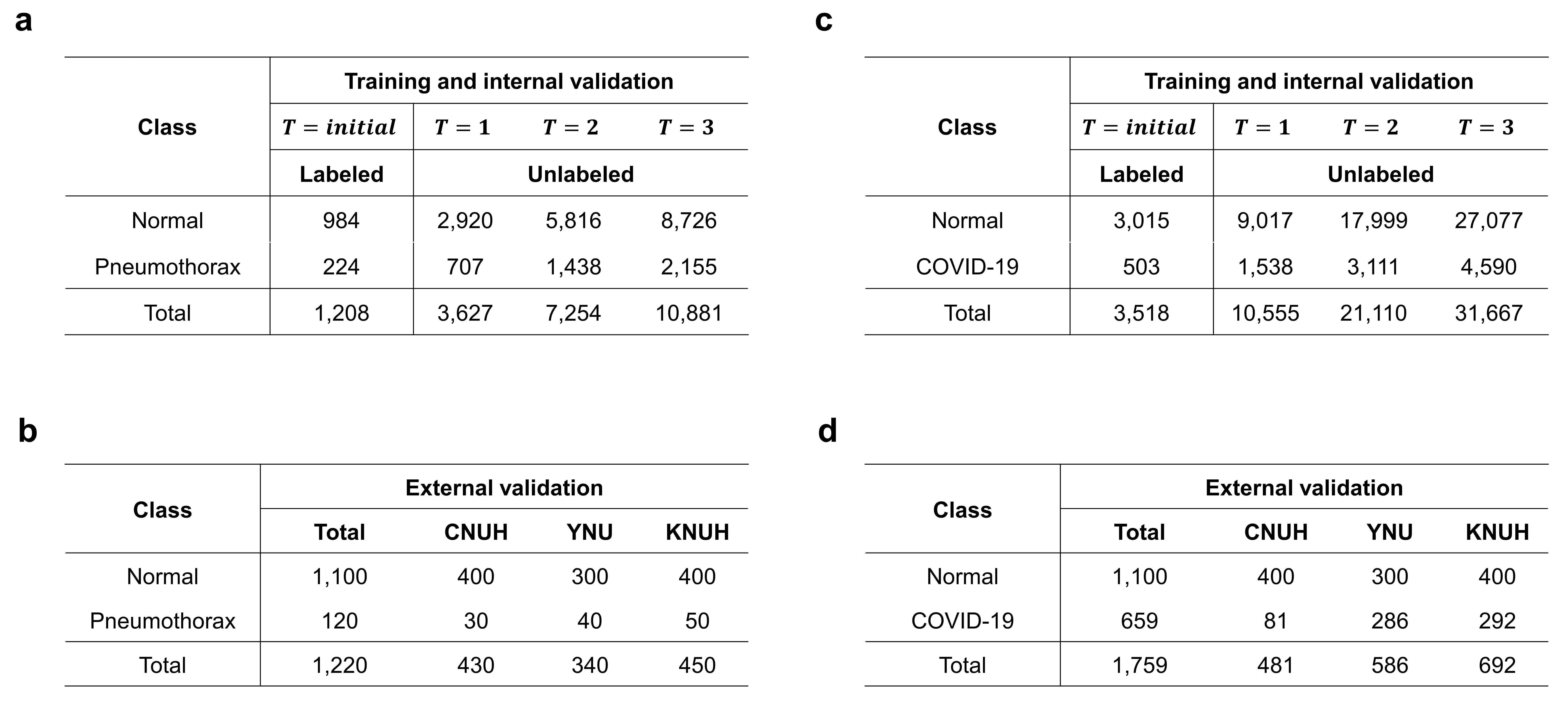, width=1.0\linewidth}}
	\caption{\bf\footnotesize 
	Data used for the experiments of two other diagnosis tasks.
(a) Data partitioning used for model development in pneumothorax diagnosis task.
(b) Data for external validation in pneumothorax diagnosis task.
(c) Data partitioning used for model development in COVID-19 diagnosis task.
(d) Data for external validation in COVID-19 diagnosis task.
}
	\label{fig3}
\end{figure}

{For pneumothorax diagnosis, we used the SIIM-ACR pneumothorax data \cite{SIIMACRP2:online} for the model development and internal validation. As it contains the CXRs and the segmentation mask for either pneumothorax and normal cases, we adopted it to be the pneumothorax diagnosis task, as the binary classification problem. Similar to the tuberculosis diagnosis task, we partitioned this data into a labeled and unlabeled subset, and the unlabeled subset was further divided into three to simulate the gradually accumulating data with time (Fig.~\ref{fig3}a and ``Details of datasets for diagnosis" section). For external validation of the trained model, we also collected the CXRs of pneumothorax patients in the three hospitals (CNUH, YNU, KNUH) (Fig.~\ref{fig3}b).} {For COVID-19 diagnosis, we utilized the two publicly available COVID-19 datasets \cite{vaya2020bimcv, signoroni2020end} for the model development and internal validation (Fig.~\ref{fig3}c and ``Details of datasets for diagnosis" section) by gradually increasing the amount of unlabeled data with increasing time similar to other tasks, while the CXRs of polymerase chain reaction (PCR) confirmed COVID-19 cases were deliberately collected for the external validation in the three hospitals (CNUH, YNU, KNUH) (Fig.~\ref{fig3}d).}

%

\subsection{Our TB diagnosis model can self-evolve with increasing unlabeled data.}
\begin{figure}[!t]
	\centering
 \centerline{\epsfig{figure=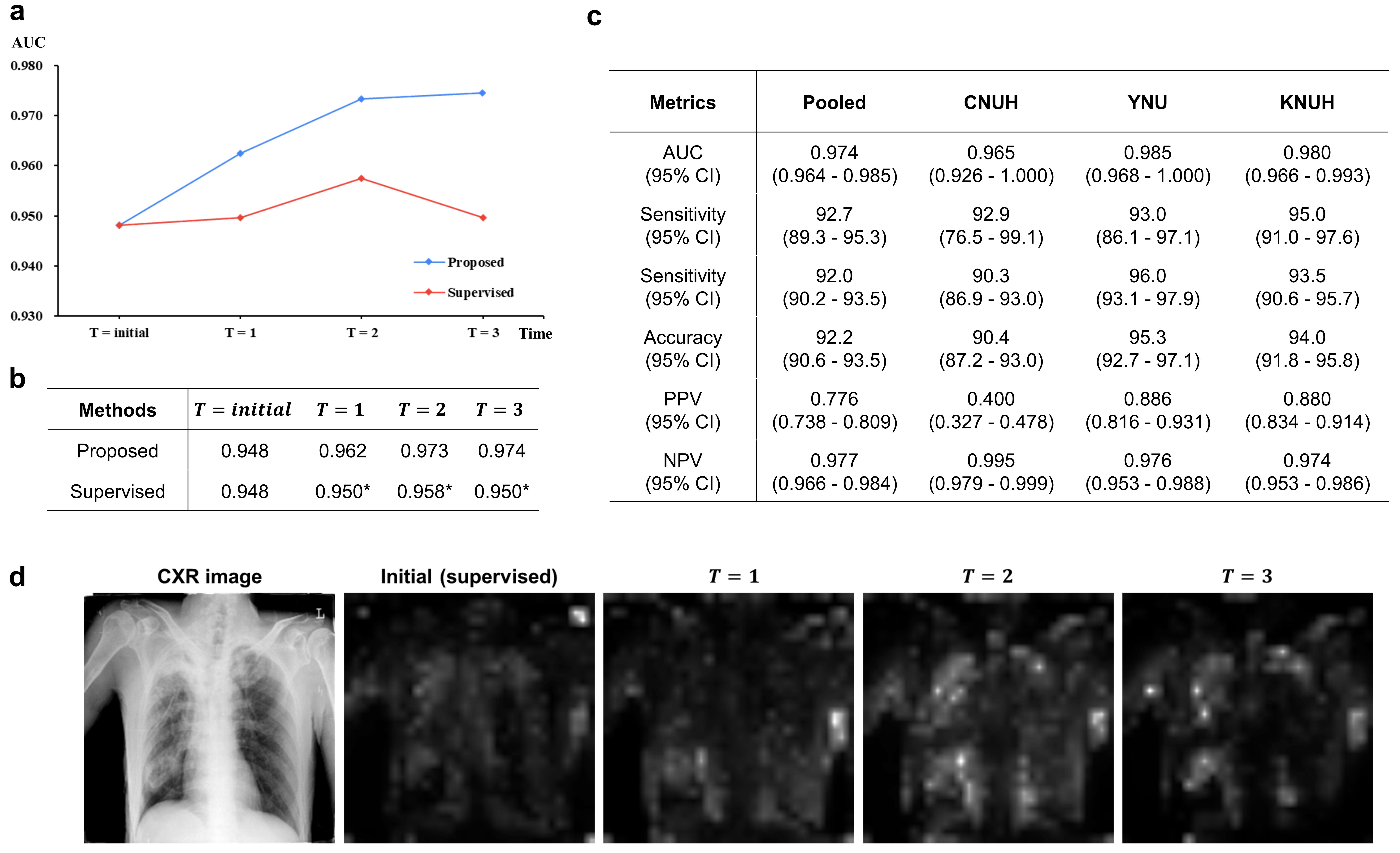, width=1.0\linewidth}}
	\caption{\bf\footnotesize 
TB diagnosis results: (a) and (b) Gradually evolving performance with the proposed framework under an increasing amount of data. Compared with the supervised model using the same amount of data with labels, the model trained with the proposed framework without labels showed even better performances.
(c) Detailed diagnostic performances of the model trained with the proposed method.
(d) Gradual attention change of the evolving model for a tuberculosis case.
For the exemplified tuberculosis case, the attention of the Vision Transformer (ViT) model gets gradually refined to better catch the target lesion and semantic structures as the model evolve with increasing $T$. $^{*}$ denotes statistically significant ($p < 0.050$) superiority of the proposed framework. AUC, area under the receiver operating characteristic curve; CI, confidence interval; PPV, positive predictive value; NPV, negative predictive value.
}
	\label{fig4}
\end{figure}

We first evaluated whether the performance of TB diagnosis can gradually be improved with the proposed framework given the increasing numbers of unlabeled data. 
As shown in Fig.~\ref{fig4}a and b, in the external validation, the performance of the model trained with the proposed framework improved as the number of unlabeled data increased, from an AUC of 0.948 to 0.974. Of note, the improved performance was even better than the supervised model trained with the same amount of data with labels, which improved to the AUC of 0.958 at $T = 2$ but decreased to 0.950 at $T = 3$, showing the sign of overfitting. In detail, the final model showed the AUCs of 0.974, 0.965, 0.985, 0.980, sensitivities of 92.7\%, 92.9\%, 93.0\%, 95.0\%, specificities of 92.0\%, 90.3\%, 96.0\%, 93.5\%, accuracies of 92.2\%, 90.4\%, 95.3\%, 94.0\% in the pooled test set and three institutions (Fig.~\ref{fig4}c), which confirmed the excellent generalization capability in clinical situation with difference devices and settings.

Not confined to the metric itself, we also observed an interesting finding that the model attention of the ViT model gets refined with increasing time $T$ (Fig.~\ref{fig4}d). As the AI model evolves with increasing time $T$, the self-attention of AI gets refined to better localize the target lesion as well as semantic structures within the given CXR image.

\begin{figure}[!t]
	\centering
 \centerline{\epsfig{figure=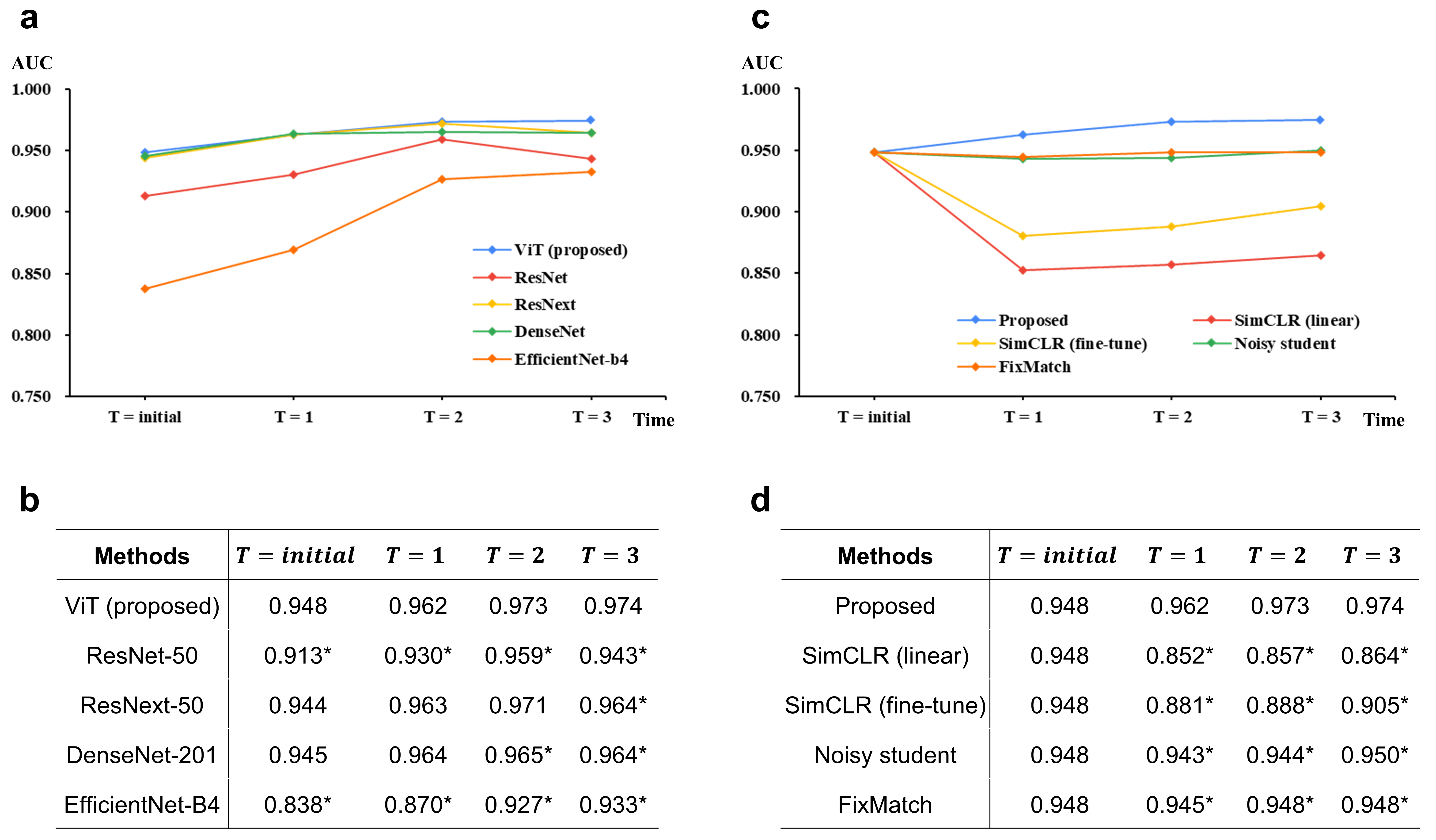, width=1.0\linewidth}}
	\caption{\bf\footnotesize 
	TB diagnosis results:
(a)(b) Compared to other convolutional neural networks (CNN)-based models, the Vision Transformer (ViT) model showed a linear increase as well as the best performance among the models.
(c) Unlike the model trained with the proposed framework, none of the existing self-supervised and semi-supervised methods showed a prominent improvement in performance.
(d) Comparison with other self-supervised and semi-supervised methods.
$^{*}$ denotes statistically significant ($p < 0.050$) superiority of the proposed framework. AUC, area under the receiver operating characteristic curve.
}
	\label{fig5}
\end{figure}

Notably, the gradual improvement of performance was prominent for the ViT model equipped with self-attention than the CNN-based models (Fig.~\ref{fig5}a and b). The ViT model showed a linear increase as well as the best performance among the models, although other CNN-based models also showed performance improvement with the proposed framework under increasing unlabeled data. In addition, the ViT model showed no sign of overfitting which was observed in some CNN-based models at later $T$.

We further evaluate whether the existing self-supervised and semi-supervised learning methods, which can also be utilized for the plenty of unlabeled data with increasing $T$, can improve the performance of the AI model gradually similar to the proposed framework (Fig.~\ref{fig5}c and d). With the same experimental settings, the existing methods presented the significant degradation of performance at $T = 1$ where the number of unlabeled data is relatively small, while the performances slightly improve with more data with increasing $T$. Even with this increase in performance, none of the existing self-supervised and semi-supervised methods showed prominent performance improvement compared with the initial model, while the model built with the proposed framework showed stably improving performance with increasing unlabeled data.

\begin{figure}[!t]
	\centering
 \centerline{\epsfig{figure=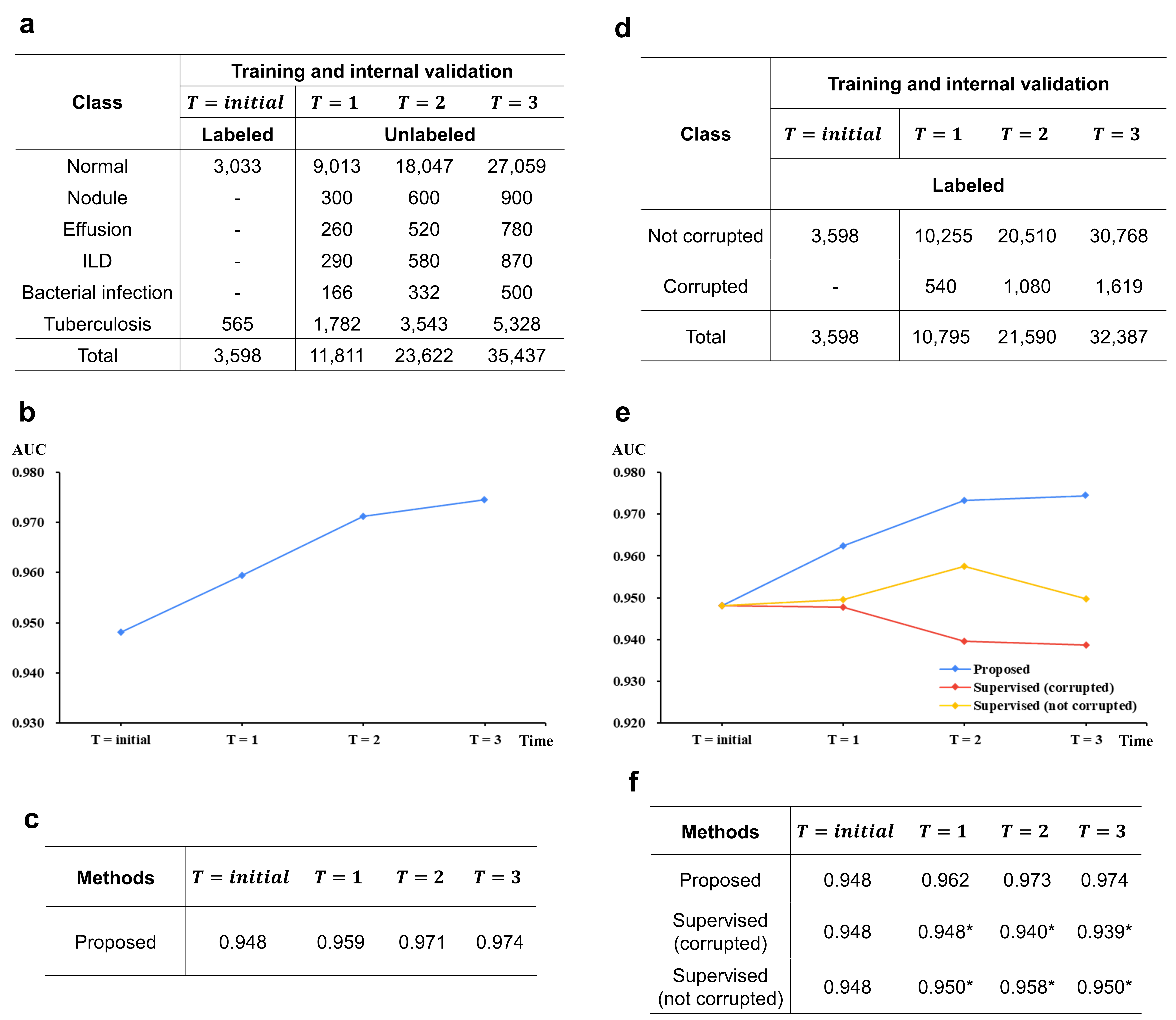, width=1.0\linewidth}}
	\caption{\bf\footnotesize 
		TB diagnosis results:
(a) Data partitioning to simulate the experiment of real-world data collection. 
(b)(c) Even after adding the unseen class data that are commonly encountered in clinics, the performance was stably improved with increasing time $T$, even though these other class data were not included for the training of the initial model.
(d) Data partitioning used to simulate the experiment of label corruption.
(e)(f) In the simulation for label corruption, the model trained with the proposed framework was not compromised, while that trained with supervised learning using corrupted labels showed significant deterioration in performance.
$^{*}$ denotes statistically significant ($p < 0.050$) superiority of the proposed framework. ILD, interstitial lung disease; AUC, area under the receiver operating characteristic curve.
}
	\label{fig6}
\end{figure}

Finally, in real clinical applications, data of totally different classes may be included when collecting the unlabeled cases, and the incorrectly labeled data may be added by the mistake of a practitioner. Therefore, we performed two experiments to verify the robustness of the proposed framework in these situations. 
First, we collected data of four other classes (nodule, effusion, interstitial lung disease, bacterial infection) that are commonly encountered in clinics from a hospital (Asan Medical Center [AMC]). These other class data were added in the same manner when increasing the number of unlabeled data over time (Fig.~\ref{fig6}a). Notably, the performance was stably improved as the same in the experiments without adding these other classes data (Fig.~\ref{fig6}b and c), suggesting the robustness of the proposed framework assuring that the AI model is not confused by these unfamiliar classes to the initial model trained only with normal and tuberculosis data. Secondly, we randomly make the label wrong with a probability of 5\% for the supervised learning and evaluated whether the performance decreased (Fig.~\ref{fig6}d). The model trained with supervised learning using the corrupted label showed significant deterioration in performance, while that with the proposed framework was not altered as it does not depend on the label for increased data (Fig.~\ref{fig6}e and f). Taken together, these results suggest the impressive reliability of the proposed framework which is required in real clinical application.


Under the hypothesis that ViT's direct attention can provide better localization than CNN's indirect attention via the {Gradient-weighted Class Activation Mapping (GradCAM)\cite{selvaraju2017grad}}, we quantified the localization performance with model attention. A total of 30 CXRs in the external validation data for TB diagnosis were selected, and manually annotated by the clinician. The predictions from model attention were generated by applying the threshold values after normalization to best localize the target lesions (0.1 for ViT and 0.6 for CNN models). As the ViT model has multiple heads to be visualized, the best performing head was selected for evaluation. The dice similarity coefficients were calculated to assess the consistency between the predictions and labels.

\textsc{\begin{figure}[!t]
	\centering
 \centerline{\epsfig{figure=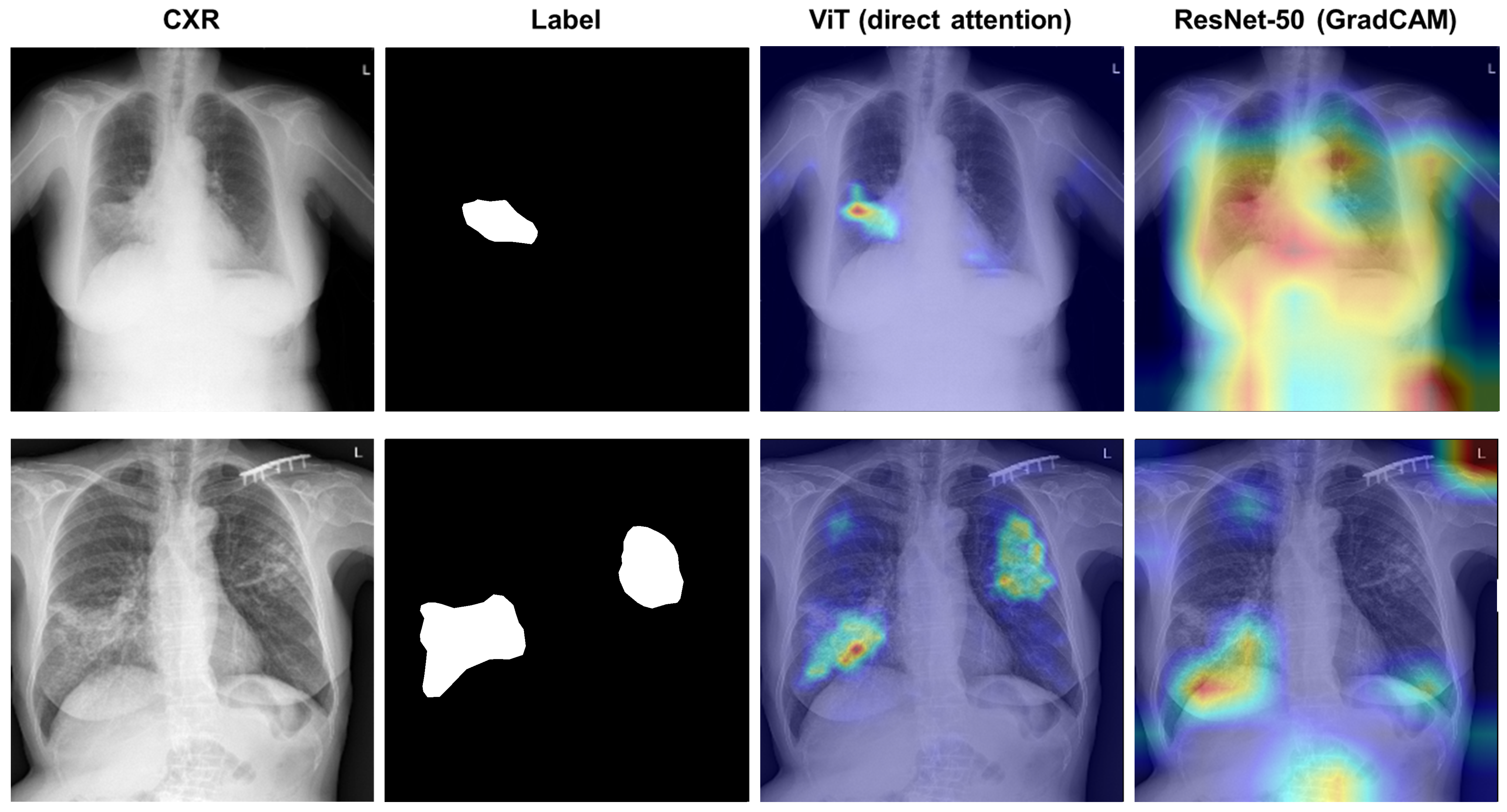, width=1.0\linewidth}}
	\caption{\bf\footnotesize 
TB diagnosis results: examples of localization with attention by Vision Transformer (ViT) and convolutional neural network (CNN)-based models.
The direct visualization of ViT attention offers better localization of the target lesion than indirect attention visualization of CNN-based models using Gradient-weighted Class Activation Mapping (GradCAM).
}
	\label{fig7}
\end{figure}}

Without any supervision during the training, the direct visualization of ViT attention offered better localization of the target lesion than the indirect attention visualization of CNN-based models using GradCAM, providing a mean dice similarity coefficient of 0.622 (standard deviation [STD] of 0.168) compared with that of 0.373 (STD of 0.259) for a CNN-based model. Of note, the indirect attention using GradCAM either attends on unimportant location (upper figure) or fails to localize the multiple lesions (lower figure) (Fig.~\ref{fig7}).

\subsection{Verifying applicability of the proposed framework in other tasks.}
\begin{figure}[!t]
	\centering
 \centerline{\epsfig{figure=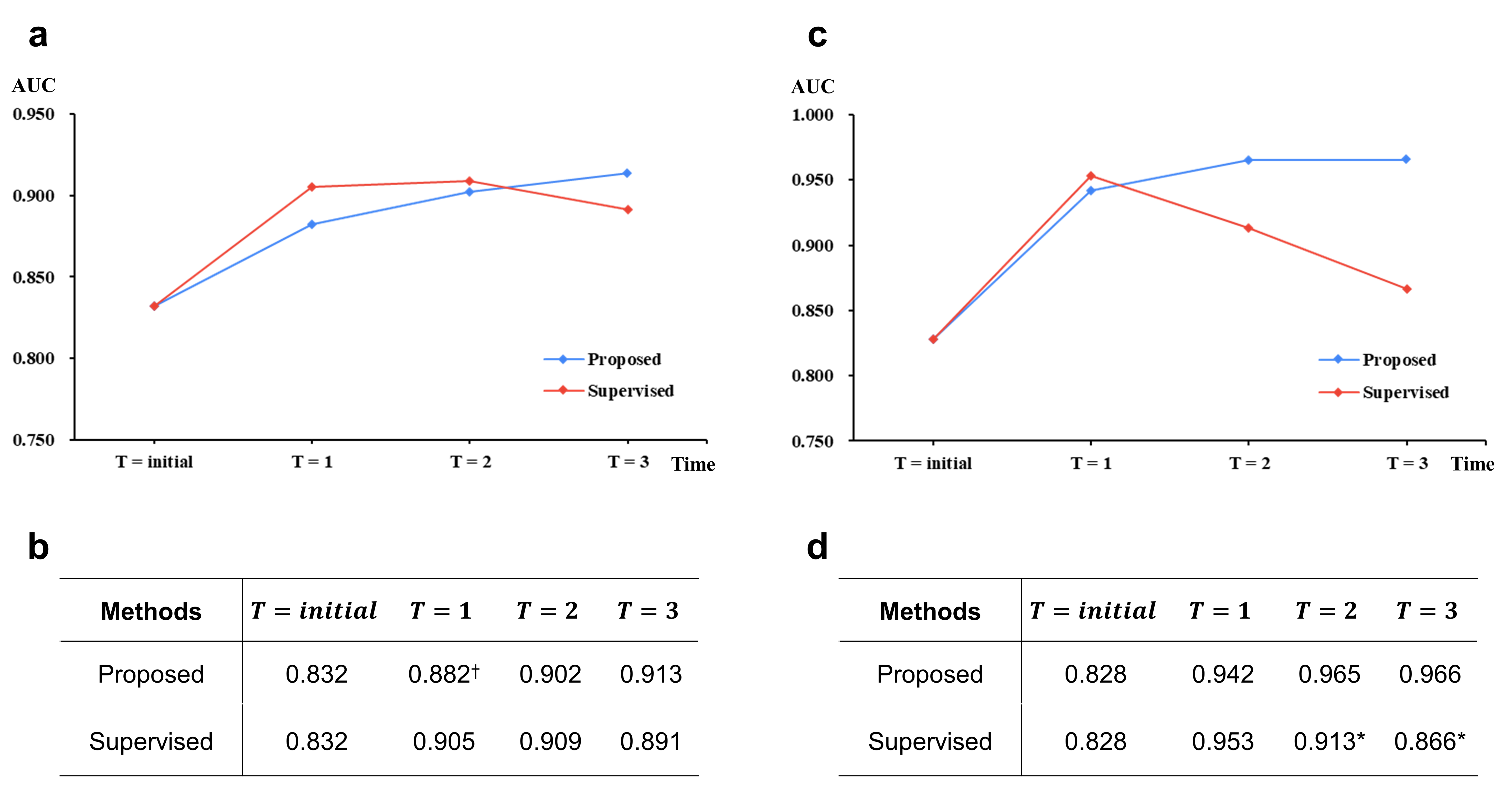, width=1.0\linewidth}}
	\caption{\bf\footnotesize 
Other CXR diagnosis results: (a)(b) When applied for pneumothorax diagnosis, the model trained with the proposed framework improved gradually over increasing times $T$, while the supervised model showed the sign of overfitting at later $T$.
$^{\dagger}$ denotes statistically significant ($p < 0.050$) superiority of the proposed framework. AUC, area under the receiver operating characteristic curve.
(c)(d) Likewise, the model trained with the proposed framework outperformed the supervised model for the COVID-19 diagnosis task, which was more prominent under the increasing amount of unlabeled data.
$^{*}$ denotes statistically significant ($p < 0.050$) superiority of the proposed framework. COVID-19, coronavirus disease 2019; AUC, area under the receiver operating characteristic curve.
}
	\label{fig8}
\end{figure}

We further analyze whether the gradual performance improvement with the proposed framework can also be observed in the CXR tasks other than tuberculosis diagnosis. First, for pneumothorax diagnosis, similar to the observation in the tuberculosis diagnosis task, the model trained with the proposed framework improved gradually over increasing times $T$ (Fig.~\ref{fig8}a and b). Notably, the performance of the model with the proposed framework was lower than the supervised one when available unlabeled data are relatively small ($T = 1$) but it ultimately outperformed the supervised model with the increased numbers of unlabeled data ($T = 3$). 
Similarly, for COVID-19 diagnosis,  the proposed framework provided the stable performance improvement over time, whereas the model trained with the same amount of labeled data showed a substantial performance drop at later $T$ in the external validation (Fig.~\ref{fig8}c and d), suggesting that overfitting to training data degraded the generalization performance of the supervised model.

\section*{Discussion}
Given the striking results of early studies that AI can keep up or even surpass the performance of the experienced clinician in various applications in medical imaging \cite{gulshan2016development, de2018clinically}, we have confronted the era of flooding AI models for medical imaging. However, these models share a common drawback that they highly depend on the quantity and quality of labels as well as the data. If the labeled corpus does not contain sufficient data points to represent the entire distribution, the resulting model can be biased and the generalization performance can unpardonably deteriorate. In the field of medical imaging, a large number of raw data is being accumulated each year without label annotation. With the supervised learning approaches, it is difficult to utilize this large corpus of unlabeled data. Therefore,  several methods based on unsupervised learning \cite{raza2021tour, ahn2019unsupervised}, self-supervised learning \cite{chen2019self} and semi-supervised learning \cite{liu2020semi} have been proposed to cope with this problem, but their performances are still sub-optimal.

{To cope with this problem,}
the proposed framework stands based on two key components: self-supervised learning and noisy self-training with knowledge distillation,
which offers stably evolving performance simply with an increasing amount of unlabeled data. The first component, in our method, is similar to that proposed in a previous work \cite{caron2021emerging}, which encourages the model to learn the task-agnostic semantic information of the image by the local-global correspondence. In our preliminary experiment, the model built only with this self-supervision attends noticeably well on the image layout, and particularly, object boundaries as shown in Supplementary Fig.~1. Secondly, the semi-supervised component enables the model to directly learn the task-specific features, the diagnosis of tuberculosis, similar to the noisy student self-training \cite{xie2020self}. Under the continuity and clustering assumption, \cite{kim2021recent}, learning with a soft-pseudo label along with student-side noise increases not only the performance but also the robustness on adversarial samples. 

Interestingly, we have found an analogy between the proposed framework and the training process of radiologists during their junior years. When a junior radiologist learns to read CXR, a common practice is to first read ``\emph{CXR}" and affirm it with ``\emph{computed tomography}" image of the same patient, which usually offers a more accurate diagnosis. This procedure is analogous to the learning process of the student in our framework in which the model learns to match the prediction of the ``\emph{noisy}" augmented image to that of the ``\emph{clean}" original image by the teacher which offers a more accurate diagnosis. In addition, it is also a common practice that the ``\emph{junior}" radiologist learns referring to the ``\emph{senior}" radiologist's reading, which is similar to the ``\emph{teacher-student}" distillation used in our framework. Finally, during the learning process, the junior radiologists occasionally refer to the ``\emph{text book}" containing small but typical cases, which prevents from being biased from recently seen atypical cases. In our framework, the ``\emph{correction step}" with the small number of initial labeled data plays a similar role.
As a result, the proposed framework, unlike the existing self-supervised and semi-supervised learning approaches, offered gradually evolving performance simply by increasing the amount of unlabeled data, with the substantial robustness to the data corruption from data of different classes or label corruption. In addition, we found in the experiments extending the application of our framework to the pneumothorax and COVID-19 diagnosis that it provides the benefit generally applicable to a variety of tasks.

Practically, our method holds great potential for {the screening of diseases like tuberculosis}, especially when applied in underprivileged areas. In the simulation of application of the model under real-world prevalence, it shows a negatively predicted portion of 72.5\% and a negative predictive value (NPV) of 0.977. That is to say, it can rule out the 72.5\% of the screened population from further evaluation by a clinician with the probability of 97.7\%, resulting in a substantial decrease of workload in resource-limited settings. In addition, the AI model can improve the performance by itself, using the proposed {DISTL} method and the iterative self-evolving framework without any further supervision by human experts. This is another important merit to be used in underprivileged areas, where plenty of data are available due to the high prevalence of diseases but the number of experts is scanty.

This study has several limitations. First, the details concerning the patient demographics and CXR characteristics were not available in some open-source data used for the training and internal validation. Second, although we simulated the robustness to unseen class data assuming real-world data collection in the experiment, it was not possible to consider all the other minor classes that can be considered in real-world data accumulation. Third, we utilized a total of 35,985 CXRs to demonstrate the benefit of the proposed framework by using them after dividing them into the small labeled and large unlabeled subsets, but the number may be insufficient to draw a firm conclusion. Further studies are warranted to verify the proposed framework in a data corpus large enough to represent the general distribution. Nevertheless, with the data-abundant but label-insufficient condition being common for medical imaging, we believe that it may offer great applicability to a broad field of medical imaging.

\section*{Methods}

\subsection{Details of datasets for pre-training.}
To pretrain the model to learn {task-relevant CXR} feature in a large corpus of CXRs, we used the CheXpert dataset~\cite{irvin2019chexpert} containing 10 common CXR classes: no finding, cardiomegaly, lung opacity, consolidation, edema, pneumonia, atelectasis, pneumothorax, pleural effusion, and support device. Among the 10 classes, five classes including lung opacity, consolidation, edema, pneumonia, and pleural effusion, considered to be related to the manifestation of infectious disease, were selected as {task-relevant CXR features}. Consequently, the model was first trained to classify these five classes with the CheXpert data. With a total of 224,316 CXRs from 65,240 subjects, 29,420 posterior-anterior (PA) and 161,427 anterior-posterior (AP) view CXRs were used after excluding the 32,387 lateral view CXRs. Thanks to this huge number of cases, the model was able to be a robust extractor for {the task-relevant CXR features}, without depending upon the variation in patients and the setting for image acquisition. As suggested in the ablation study of pre-training (see Supplementary Fig.~2), this pre-training step has brought us a substantial increase in performance and is one of the key components of our model.

\subsection{Details of datasets for diagnosis.}
{
First, for the tuberculosis diagnosis, we used both public and institutional data sets for the model development and internal validation. Specifically, data deliberately collected from a hospital (Asan Medical Center [AMC]) as well as publicly available data (National Institutes of Health [NIH] \cite{TBPortal59:online}, Valencian Region Medical ImageBank [BIMCV] \cite{vaya2020bimcv}, CheXpert \cite{irvin2019chexpert}, India \cite{TBXpredi94:online}, Montgomery, Shenzen \cite{jaeger2014two}, Belarus \cite{tbcnnbel14:online}, PAthology Detection in Chest radiograph [PADChest] \cite{bustos2020padchest}, Tuberculosis X-ray 11K [TBX 11K] \cite{liu2020rethinking}) were integrated (Supplementary Fig.~3), to have a total of 35,985 CXRs containing 5,893 tuberculosis and 30,092 normal cases.}
{For pneumothorax diagnosis, the SIIM-ACR pneumothorax data \cite{SIIMACRP2:online} were utilized for the model development and internal validation. The original SIIM-ACR pneumothorax data contains 12,089 CXR images with or without corresponding masks for the 2,379 pneumothorax or 9,710 normal cases. Therefore, we adopted the problem into a binary classification for the diagnosis of pneumothorax by defining CXR as a pneumothorax positive case if a segmentation mask contains a positive value and as a negative case if not.}
{Finally, for COVID-19 diagnosis, two publicly available open-sourced datasets were used for the model development and internal validation \cite{vaya2020bimcv, signoroni2020end}. As these two datasets contain a small number of normal CXRs, we also utilized the normal CXRs in the tuberculosis diagnosis task as the normal cases, finally yielding a total of 35,185 CXRs consisting of 5093 COVID-19 and 30,092 normal cases.}

{For all the three CXR diagnosis tasks, data collected from three hospitals (Chonnam National University Hospital [CNUH], Yeungnam University Hospital [YNU], and Kyungpook National University Hospital [KNUH], labeled by board-certified radiologists, were used to validate the generalization capability of the model for different devices and image acquisition settings. In detail, 328 tuberculosis and 1,100 normal CXRs for the tuberculosis diagnosis task, 120 pneumothorax and 1,100 normal CXRs for the pneumothorax diagnosis task, and 120 COVID-19 and 1,100 normal CXRs are evaluated for the external validation of each task.
}

\subsection{Details of Implementation.}
The CXR images underwent preprocessing including histogram equalization, Gaussian blurring, and normalization, and finally resized to $256 \times 256$. 
As the backbone part $f$ of the network, we used the ViT small model (12 layers and six heads) with the patch size of $256 \times 256$, and the CNN-based models (ResNet-50, DenseNet-201, ResNext-50, and EfficientNet-B4) were used for comparison. For the classification and projection heads, the three-layered multi-layer perceptron (MLP) was utilized. 
For pre-training, an Adam optimizer was used with a learning rate of 0.0001. The model was pre-trained for 5 epochs with a step decay scheduler, with a batch size of 16. Weak data augmentation including random flipping, rotation, translation as suggested in \cite{ye2020weakly} were performed to increase the data variability during the pre-training. As the loss functions, binary cross-entropy 
 (BCE) losses were used for each class label.
 For supervised training of the initial model and the iterative training of the models with DISTL, Adam W optimizer \cite{loshchilov2017decoupled} were used along with cosine decay scheduler with a maximum learning rate of 0.00005. The model was trained for 5 epochs with one epoch for the warm-up step. The correction step is performed per 500 updates. Similar to pre-training, and a batch size of 16 was used and weak augmentation was performed during the training. The difference between the training of initial model training and iterative training with DISTL is that multi-crop strategies were used with the scale range of 0.75-1 for global crop and 0.2-0.6 for local crops to yield multiple different views for a given image. The cross-entropy loss was used as the loss function for both the classification and the self-supervising losses.
All experiments including preprocessing, model development, and evaluation were performed using Python version 3.8 and Pytorch library version 1.8 on NVIDIA Quadro 6000, GeForce RTX 3090, and RTX 2080 Ti.

\subsection{Details of evaluation.}
For evaluation of overall model performances, the three independent test sets were pooled and then used to evaluate the overall performance of the model, while the model performances in three individual test sets were reported separately.
The area under the receiver operating characteristics curve (AUC) was used as the primary evaluation metric, and the sensitivity, specificity, and accuracy were also calculated to meet the pre-defined sensitivity value $\geq 80 \%$ by adjusting the thresholds, if possible. 
To statistically compare the proposed method with others, the DeLong test was performed to estimate 95\% confidence intervals (CIs) and p-values. Statistically significant differences were defined as p $<$ 0.05.

Direct visualization of the model attention is another merit of the ViT model. Similar to the approach introduced in a self-supervised learning approach for ViT \cite{caron2021emerging}, we used the attention weights of multi-head in the last layer of the Transformer encoder to visualize attention. For comparison, the model attention was visualized indirectly with the Grad-CAM \cite{selvaraju2017grad} that generates the model attention with the linear combination determined by the gradients of the output with regard to the last layer feature map, for CNN-based model.

\subsection{Ethic committee approval.}
The four hospital data deliberately collected for this study were ethically approved by the Institutional Review Board of each hospital and the requirement for informed consent was waived.

 \begin{addendum}
{\color{black} \item[Correspondence] Correspondence and requests for materials should be addressed to Jong Chul Ye.~(email: jong.ye@kaist.ac.kr).}
 \item  This research was supported by the National Research Foundation (NRF) of Korea under Grant NRF-2020R1A2B5B03001980.
{
\item[Author Contributions] S.P. performed all experiments, wrote the extended code, and prepared the manuscript. G.K and Y.O. contributed in data preprocessing. J.B.S, S.M.L., J.H.K, S.M, and J.K.L collected and labeled data. C.M.P advised the project in conception. J.C.Y. supervised the project in conception and discussion, and prepared the manuscript.}
 \item[Competing Interests] 
The authors declare that they have no competing financial interests.
 {
  \item[Data Availability] 
Part of CXRs are compiled from publicly available open-source data repositories. The CheXpert repository is available at \href{https://stanfordmlgroup.github.io/competitions/chexpert/}{https://stanfordmlgroup.github.io/competitions/chexpert/}, The BIMCV repository is available at \href{https://github.com/BIMCV-CSUSP/BIMCV-COVID-19}{https://github.com/BIMCV-CSUSP/BIMCV-COVID-19}. The India tuberculosis repository can be found at \href{https://www.kaggle.com/raddar/chest-xrays-tuberculosis-from-india}{https://www.kaggle.com/raddar/chest-xrays-tuberculosis-from-india}. Montgomery and Shenzen data can be requested via the contact on the follwing webpage \href{https://openi.nlm.nih.gov/}{https://openi.nlm.nih.gov/}. Belarus tuberculosis repository is available at \href{https://github.com/frapa/tbcnn/tree/master/belarus}{https://github.com/frapa/tbcnn/tree/master/belarus}. The PADChest repository is available at \href{https://github.com/auriml/Rx-thorax-automatic-captioning}{https://github.com/auriml/Rx-thorax-automatic-captioning}. The TBX 11K repository can be accessed at \href{https://www.kaggle.com/usmanshams/tbx-11}{https://www.kaggle.com/usmanshams/tbx-11}. SIIM-ACR Pneumohtorax Segmentation dataset is available at the following repository \href{https://www.kaggle.com/c/siim-acr-pneumothorax-segmentation}{https://www.kaggle.com/c/siim-acr-pneumothorax-segmentation}. Brixia COVID-19 data repository can be found at \href{https://brixia.github.io/}{https://brixia.github.io/}.
  }
  {
  \item[Code Availability] 
The code is available at the following github repository. \href{https://github.com/depecher/}{https://github.com/depecher/}
  }
\end{addendum}

\section*{References}
\bibliographystyle{naturemag}
\bibliography{ref}

\end{document}